\documentstyle[11pt]{article}
\input{psfig.sty}
\def\ref#1{\noindent\hangindent=24.0pt\hangafter=1{#1}\par}

\def\la{\hbox{\rlap{\raise.3ex\hbox{$<$}}\lower.8ex\hbox{$\sim$}\ }}
\def\ga{\hbox{\rlap{\raise.3ex\hbox{$>$}}\lower.8ex\hbox{$\sim$}\ }}

\def\ni{\noindent}

\def\pb{\parbox}
\def\pb5{\parbox[t]{4.5cm}}

\def\cm2sec{\ cm$^{2}$-sec}

\def\arcmin{$\,^\prime$~}

\def\deg{$^{\circ}$~}

\def\etal{{\it et al.\/}}

\def\H2S1{H$_2$ $S$(1)}

\def\etal{{\it et al.}}

\def\ni{\noindent}

\begin{document}

\large

\ni
{\bf BALLOON-BORNE HARD X-RAY IMAGING AND FUTURE SURVEYS}
\bigskip

\bigskip

\normalsize

\ni
Jonathan E. Grindlay
\bigskip

\ni
{\it Harvard-Smithsonian Center for Astrophysics, 60 Garden St., 
Cambridge, MA 02138, USA}

\bigskip

\ni
ABSTRACT
\medskip

\ni
Several payloads for hard  X-ray (20-600 keV) imaging 
with coded aperture telescopes have been developed for balloon 
flight observations of cosmic x-ray sources.  We briefly review the 
characteristics of these, particularly the EXITE2 system. 
The recent NASA program to develop an extended long 
duration (100d) balloon flight capability employing 
super-pressure balloons would allow a qualitatively new 
hard x-ray imaging experiment: the  
Energetic X-ray Imaging  Survey Telescope--Long 
Integration Time Experiment  (EXIST-LITE). The longer 
continuous viewing  times (per source) available from an 
LDB platform than from low earth orbit would enable both 
surveys and objectives complementary to the EXIST mission 
proposed for a MIDEX satellite.   
We summarize the scientific objectives of EXIST-LITE, a possible 
instrumentation approach incorporating a large area array of 
Cd-Zn-Te (CZT) detectors, and our program 
for the development and balloon flight testing of relatively thick 
(5mm) CZT detector arrays.  
\bigskip

\ni
INTRODUCTION
\medskip

\ni
Over the past decade several balloon-borne hard x-ray imaging 
telescopes have been developed and flown. These have enabled 
the first demonstration of coded aperture imaging 
of cosmic x-ray sources  at hard x-ray 
energies (\ga30 keV) with position-sensitive scintillation detectors 
and coded masks. The first-generation hard x-ray ($\sim$20-300 keV) 
coded aperture imagers, GRIP1 and EXITE1 (cf. Table 1 below) 
preceeded the first satellite-borne imaging telescope, SIGMA 
(cf. Paul et al 1991), which established the rich variety of 
hard x-ray sources available for study from a long-exposure mission. 
\medskip

\ni
In this paper we provide a brief overview of 
the several balloon-borne hard x-ray imagers developed  
including our EXITE detectors, particularly the 
current EXITE2 system. These have all been small (\la5\deg) 
or moderate ($\sim$15\deg)  
field of view instruments with emphasis on study of individual 
sources. With the planning now for  possible long-duration balloon (LDB) 
flights at mid-latitudes, it is now possible to consider future 
balloon-borne wide-field imagers which could conduct surveys 
and studies of many sources simultaneously. Such a wide-field, long-exposure 
survey telescope could be developed and flown on LDB missions as 
a precursor, or even as an extended follow-on, to a full satellite mission.
\medskip

\ni
The Energetic X-ray Imaging Survey Telescope (EXIST) was proposed in
December 1994 as a New Mission Concept for 
a satellite-borne mission. It would conduct the first
imaging survey of the sky at hard x-ray energies (10-600 keV) with a
sensitivity some 100$\times$ greater than the only previous all-sky
survey carried out by HEAO-A4 experiment in 1978-80 
(Levine et al 1984). An overall description of the initial EXIST 
concept is given by Grindlay et al (1995) (and on the Web site 
{\it http://hea-www.harvard.edu/EXIST/EXIST.html}). 
EXIST was accepted for study as  a New Mission Concept, 
and has been developed
extensively  in the course of preparation and submission of
a successful ``Step 1" and solicited ``Step 2'' 
proposal for the MIDEX program. Although EXIST 
was not selected (June 1996) for flight, the need for such a 
satellite mission is just as acute and the Concept Study is continuing 
(cf. Grindlay et al 1997) for a future MIDEX proposal submission. 
\medskip

\ni
Some of the objectives of EXIST could be met by a 
balloon-borne version {\it provided that the recently announced NASA 
program to develop a 100d capability for long duration balloon (LDB) flights 
is realized}. In this paper we consider the application of such a 100d  LDB 
(hereafter, 100dLDB) program to develop a version of 
EXIST which would allow for Long Integration Time Experiments on 
a variety of sources. The EXIST-LITE concept for 100dLDB fights is 
described here and contrasted with the full satellite-borne EXIST mission. 
EXIST-LITE could either preceed EXIST (as a development step, and probably 
as a smaller total detector/telescope combination) or as a follow-on mission 
to the (proposed) two year EXIST (MIDEX) mission which would enable 
continuing surveys, monitoring and study of individual sources.
\bigskip

\bigskip

\ni
BALLOON-BORNE HARD X-RAY IMAGERS
\medskip

\ni
At energies above $\sim$15 keV, it is increasingly difficult to image x-rays 
with grazing incidence optics; at energies \ga100 keV, where even graded 
multi-layer coatings lose their effectiveness, it is essentially 
impossible. Thus coded aperture imaging (cf. Caroli et al 1987 for a review),  
in which the shadow of an aperture mask with $\sim$50\% open hole fraction 
is measured by a position-sensitive detector, has enabled the development 
of several telescopes for hard x-ray (20-600 keV) and soft 
$\gamma$-ray ($\sim$100 keV - 1 MeV) imaging studies of cosmic sources. 
These imagers have employed uniformly redundant array (URA) coded masks and 
scintillation detectors (NaI(Tl) or CsI(Na)), or gas-filled 
proportional counters, for which the required position 
sensitive readout was achieved by a variety of techniques. 
Table 1  lists the key characteristics of the balloon-borne hard 
x-ray imagers in the approximate order of their 
development or flight. In addition a soft $\gamma$-ray ($\sim$200 keV
- 9 MeV) coded aperture telescope with wide field ($\sim$20\deg) and 
moderate resolution ($\sim$5\deg) was developed and flown by the 
UNH group (McConnell et al 1987). 

\medskip

\pagestyle{empty}
\setcounter{secnumdepth}{5}

\centerline{Table 1: Summary of Balloon-borne Hard X-ray Imagers}
\vskip 13pt

\begin{center}
\begin{tabular}{lrllll}
Instr. &      En. range (keV) & Detector  and  Readout&  FOV &     Ang. res.&   Refs.\\ \hline
&&&&& \\
GRIP1  &    35-1500     &        NaI; Anger camera    &    15\deg  &  2\deg  &       1,2\\
EXITE1 &   20-300       &        NaI; image intens.   &    3.5\deg &  24\arcmin&  3,4\\
GRATIS &   20-150       &        CsI; pos.-sens. PMTs &   1.5\deg &  2\arcmin  &  5\\
GRIP2  &     35-1000    &          NaI/CsI; Anger cam. &   15\deg  &  33\arcmin       &  6\\
EXITE2 &    20-600      &         NaI/CsI; Anger cam.  &   4.5\deg &  22\arcmin & 7,8\\
MIXE2	&   20-100      &        Microstrip gas counter &  1.8\deg &  7\arcmin  & 9\\

\end{tabular}
\end{center}
\vskip 3pt

\small
\noindent
References:\\
\noindent 
1. Cook, W., Finger, M. and Prince, T. 1985, {\it IEEE Trans. Nuc. Sci.}, 
{\bf NS-32},  129.\\
\noindent 
2. Cook, W.R. et al 1991, {\it ApJ}, {\bf 372}, L75.\\
\noindent 
3. Grindlay, J., Garcia, M., Burg, R. and Murray, S. 1986, 
{\it IEEE Trams. Nuc. Sci.}, {\bf NS-33}, 750.\\
\noindent 
4. Covault, C., Grindlay, J. and Manandhar, R. 1992, {\it ApJ}, {\bf 388}, 65.\\
\noindent 
5. Harrison, F., Hailey, C.J., Kahn, S.M. and Ziock, K.P. 1989, 
{\it Proc. SPIE}, {\bf 1159}, 36,.\\
\noindent 
6. Schindler, S. et al 1997, {\it NIM}, {\bf A384}, 425.\\
\noindent 
7. Manandhar, R.  et al 1993, {\it Proc. SPIE}, {\bf 2006}, 200.\\
\noindent 
8. Lum, K.  et al 1997, {\it NIM}, in press.\\
\ni
9. Ramsey, B.D. et al 1997, {\it NIM}, {\bf A383}, 424.\\
\normalsize

\medskip

\ni
\underline{EXITE Telescopes}
\medskip

\ni
We developed and flew a first-generation hard x-ray balloon-borne 
imager, the Energetic X-ray Imaging Telescope Experiment (EXITE) 
in 1986-90. The EXITE1 detector (cf. Table 1) was a 34cm diameter
(round) $\times$ 0.6cm thick NaI(Tl) scintillator optically 
coupled to a large area image intensifier with electron-reducing 
optics and PIN-diode position-sensitive readout. This provided 
a relatively simple readout and good position senstivity 
(FWHM = 14mm/E$_{keV}^{0.5}$. Graded passive shields 
and cosmic ray  anti-coincidence shields reduced measured 
background levels to F(100 keV) $\sim$ 6 $\times 10^{-4}$ counts/
cm$^2$-sec-keV (for the geomagnetic latitude of 
Alice Springs, Australia) yielding a 3$\sigma$ sensitivity 
of $\sim$100 mCrab for a 3 hour observation (Covault 1991).  
\medskip

\ni
The second-generation EXITE 
detector and telescope, EXITE2 (cf. Table 1), 
is a 40cm $\times$ 40cm phoswich scintillation 
detector (1cm-NaI/2cm-CsI) read out 
by a 7 $\times$ 7 array of close-packed (square) PMTs. The 
detector area 1300 cm$^2$ (NaI), or 1600 cm$^2$ (NaI + 2cm surrounding 
CsI guard ring), is thus $\sim2\times$ the geometric area of 
the EXITE1 detector. The phoswich discrimination allows a 
background (at 100 keV) reduction by a factor of 2-3 over 
the EXITE1 detector, so that allowing for the 
area and background factors,  the sensitivity (per unit time) 
should be a factor $\sim$2 better. Both the energy 
and spatial resolution are also each improved by factors of 
$\sim$1.4, so that improved imaging and spectra are possible. 
An engineering test flight was conducted in June 1993 (Lum et al
1994) and led to improved cosmic ray rejection electronics and a 
completed on-board recording and computer processing system 
as well improved aspect system. 
Despite campaigns to launch the payload for its first science 
flight in both May and September-October 1996, winds and the 
launch queue did not allow even an attempt; the payload is 
now awaiting a May 1997 launch.
\bigskip

\ni
BALLOON-BORNE HARD X-RAY SURVEY: EXIST-LITE
\medskip

\ni
Until now all of the telescopes in the development of balloon-borne 
(and subsequent spaceflight) hard x-ray imaging have been 
relatively narrow field of view (FOV \la5-30\deg, FWHM; cf. Table 1) 
for pointed observations 
of individual fields. A survey telescope at hard x-ray energies can 
be constructed as a coded aperture telescope with a field of 
view of up to $\sim$45\deg without significant  projection 
effects or collimation by the mask aperture, assumed planar. 
Such a telescope 
would execute a continuous scan rather than fixed target pointing 
(although pointings are also possible and would be conducted) 
to cover a maximum sky fraction in minimum time and would 
be more sensitive than a scanning grazing incidence (multi-layer) 
telescope of comparable (or even larger) physical size though smaller 
effective area and field of view. For example, it can easily be shown 
that a multi-layer telescope with (currently ambitious) parameters of 
FOV $\sim$ 10\arcmin and effective area A$_{eff} \sim$ 500cm$^2$ 
in a scanning (ROSAT-like) satellite mission would be a factor of 
$\sim$10 {\it less} sensitive than a wide-field (FOV $\sim$ 20-40\deg) 
scanning coded aperture telescope with A$_{eff} \sim$ 5000 cm$^2$ 
in the 30-100 keV band (the multilayer telescope could achieve comparable 
sensitivity in the 10-30 keV band due to bright source and diffuse 
background contributions to the wide-field coded aperture imager). 
In addition, the wide-field coded aperture imager allows the 
survey to extend up to the poorly explored 100-600 keV range,
totally unaccessible to focussing optics (except for Bragg
concentrators, which can work only in a very narrow energy band). 
\medskip

\ni
The EXIST-LITE concept would (as with the EXIST for MIDEX) 
incorporate two wide-field telescopes. However because of the 
attenuation of the residual overlying atmosphere (assumed 
to be 4 g/cm$^2$, averaged over the wide FOV), the survey would 
have little response below $\sim$25 keV. Thus the background 
contributions from the cosmic diffuse flux and from 
bright galactic sources 
will be less (at $\sim$30 keV) than for MIDEX (though this is 
partially offset by atmospheric background) and the 
FOV can be even larger. Thus the separate low-energy ($\sim$10-30 keV) 
1-D collimators ($\sim$3.5\deg) can be eliminated . 
Each of the two telescopes for EXIST-LITE 
would have FOV = 45\deg $\times$ 45\deg, for a 
combined FOV =  45\deg $\times$ 90\deg, and each would (ideally) have 
total detector area of 2500 cm$^2$. 
The schematic layout of the two telescopes is shown in Figure 1. 

\vspace*{0.1in}
\hspace*{1.in}
\psfig{figure=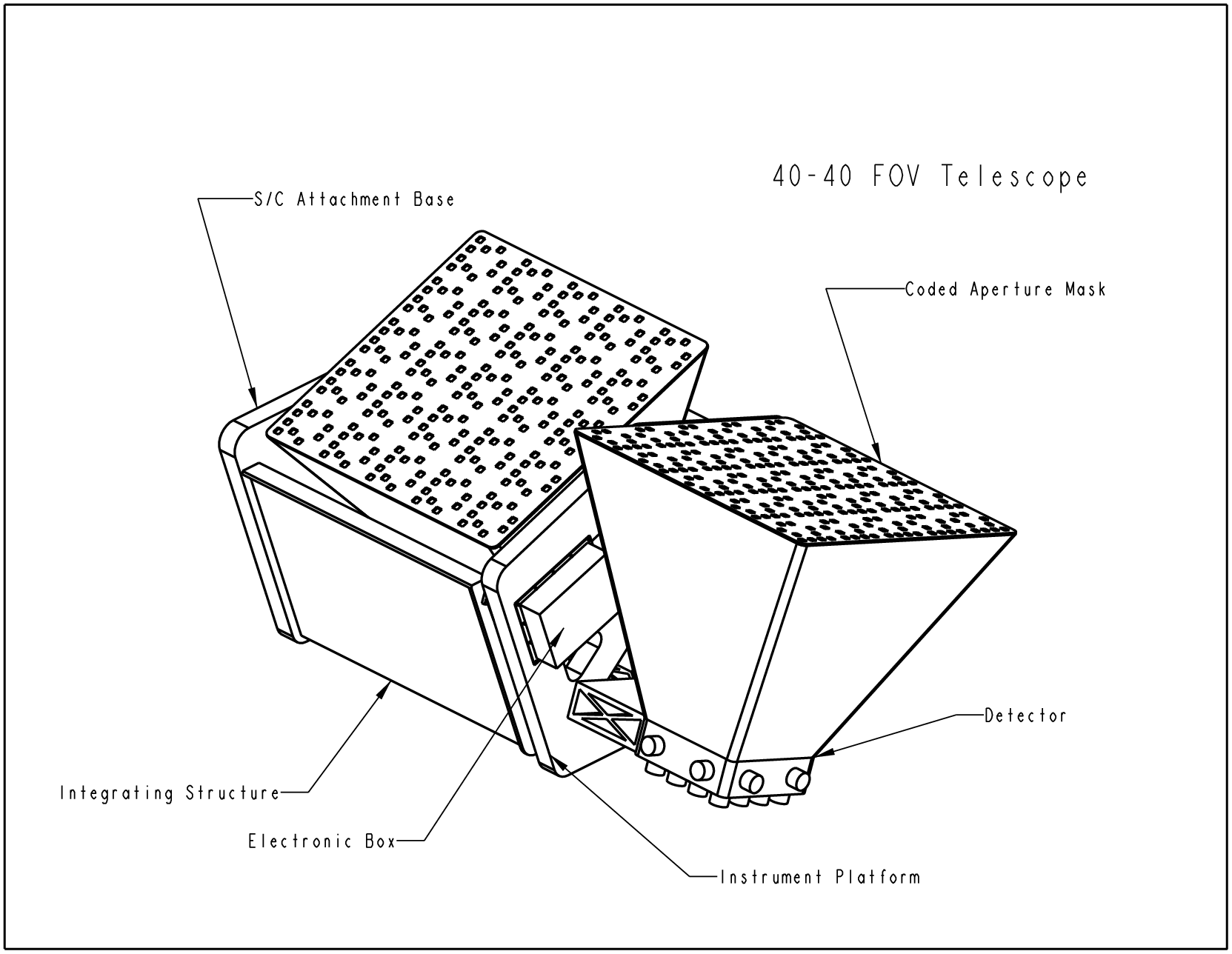,height= 3.7in,width=3.7in,angle=-90}

\vspace*{-0.6in}

\ni
Fig. 1. Schematic layout of coded aperture telescopes for EXIST or EXIST-LITE. 
\medskip

\ni
The center of the combined FOV would 
be fixed-pointed at the local zenith, and the long axis of the
combined FOV  
would be maintained north-south with a gondola pointing system, which 
would allow inertial pointing as well for a few selected priority targets 
during the survey (e.g. the M31 galaxy for a GRB survey). The sky then 
drifts east-west across the narrow (45\deg) dimension of  the FOV, giving 
a minimum exposure time each day for any given source 
with declination $\delta$ of 3h ($\delta$ = 0\deg) to 
\ga12h ($\delta$ \ga 67.5\deg). This is, of course, a continuous exposure rather 
than the interrupted exposure segments ($\sim$15-30 min) achieved in each of 
(typically) 10 orbits (non-SAA) each day for the MIDEX mission. Thus 
the key difference for EXIST-LITE is the long(er) continuous integration time 
experiments possible.
\medskip

\ni
Because of its very large FOV and large area detectors with 
high intrinsic resolution (both spatial and spectral), EXIST-LITE  
could approach the unprecedented all-sky sensitivity levels for 
the full EXIST mission, which are 
shown in  Figure 2. 
The sensitivity plots are for EXIST for its proposed 
9-month all-sky survey (followed by a pointed mission 
phase), which would allow total integration times of 
$\sim$10$^6$ sec for any source. Comparable or significantly 
greater total exposures for EXIST-LITE 
could be achieved for half the sky in a single 100dLDB flight 
since any source would be observed for \ga300-1200 hours or \ga1.1-4.4 
$\times$ 10$^6$ sec. 
Thus with just two such flights (ideally at latitudes 
$\pm$45\deg), the whole sky can be covered with exposure totals  
from $\sim1 - 4\times$ that for EXIST. For total exposures 
on a given source the same as EXIST, the sensitivity 
for EXIST-LITE would be reduced from Figure 2   
by a factor of $\sim$2-3 due to absorption/scattering in the residual 
overburden atmosphere ($\sim$4 g/cm$^2$ averaged over 
the large FOV), but this may be compensated in part 
by the lower and more stable background than encountered 
in low earth orbit (unless the orbit were equatorial). A 
detailed study of backgrounds and sensitivities for 
EXIST-LITE  will be carried out.

\hspace*{1.1in}
\psfig{figure=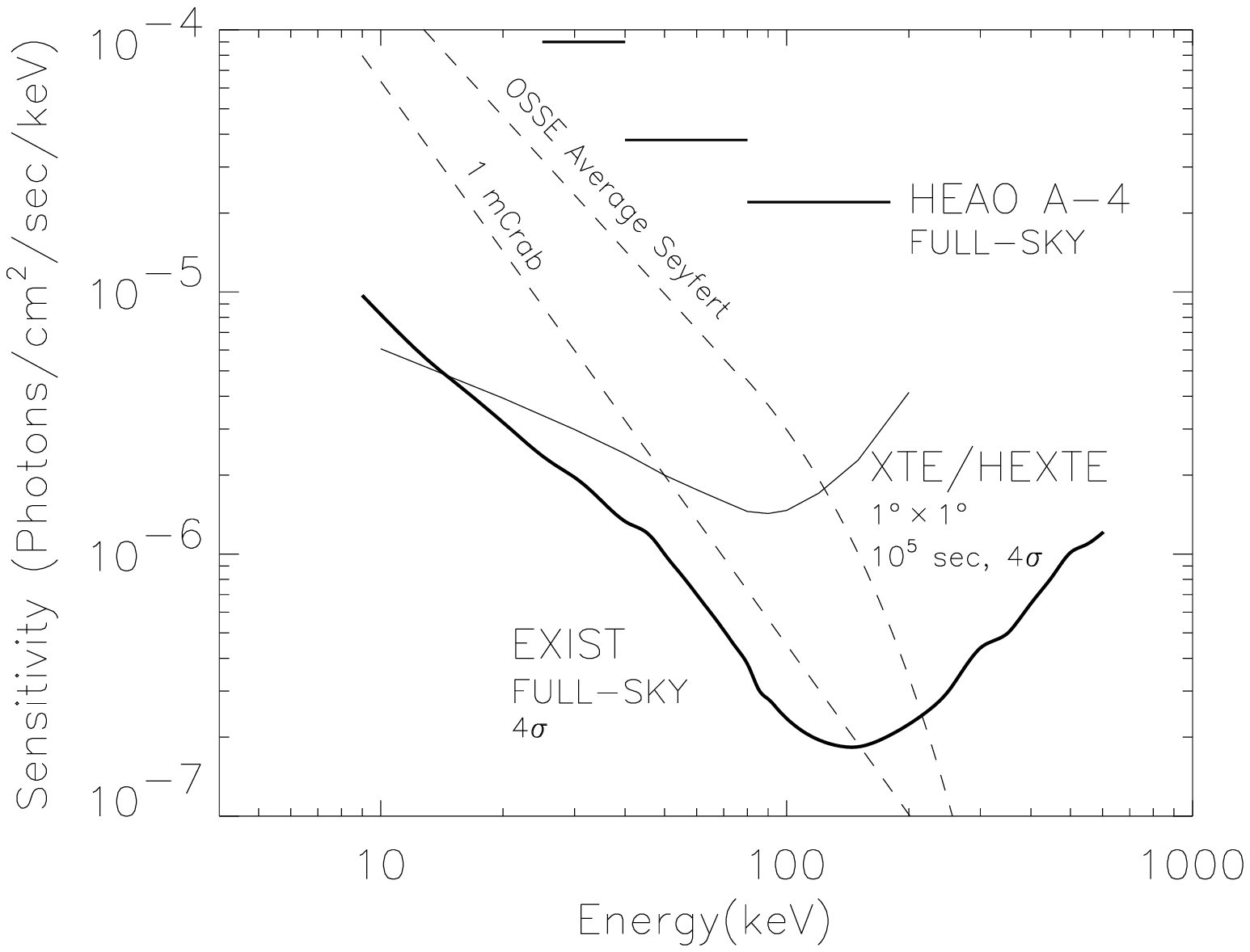,height= 3.2in,width=4.5in,angle=0}

\vspace*{-0.1in}

\hspace*{1.1in}
\psfig{figure=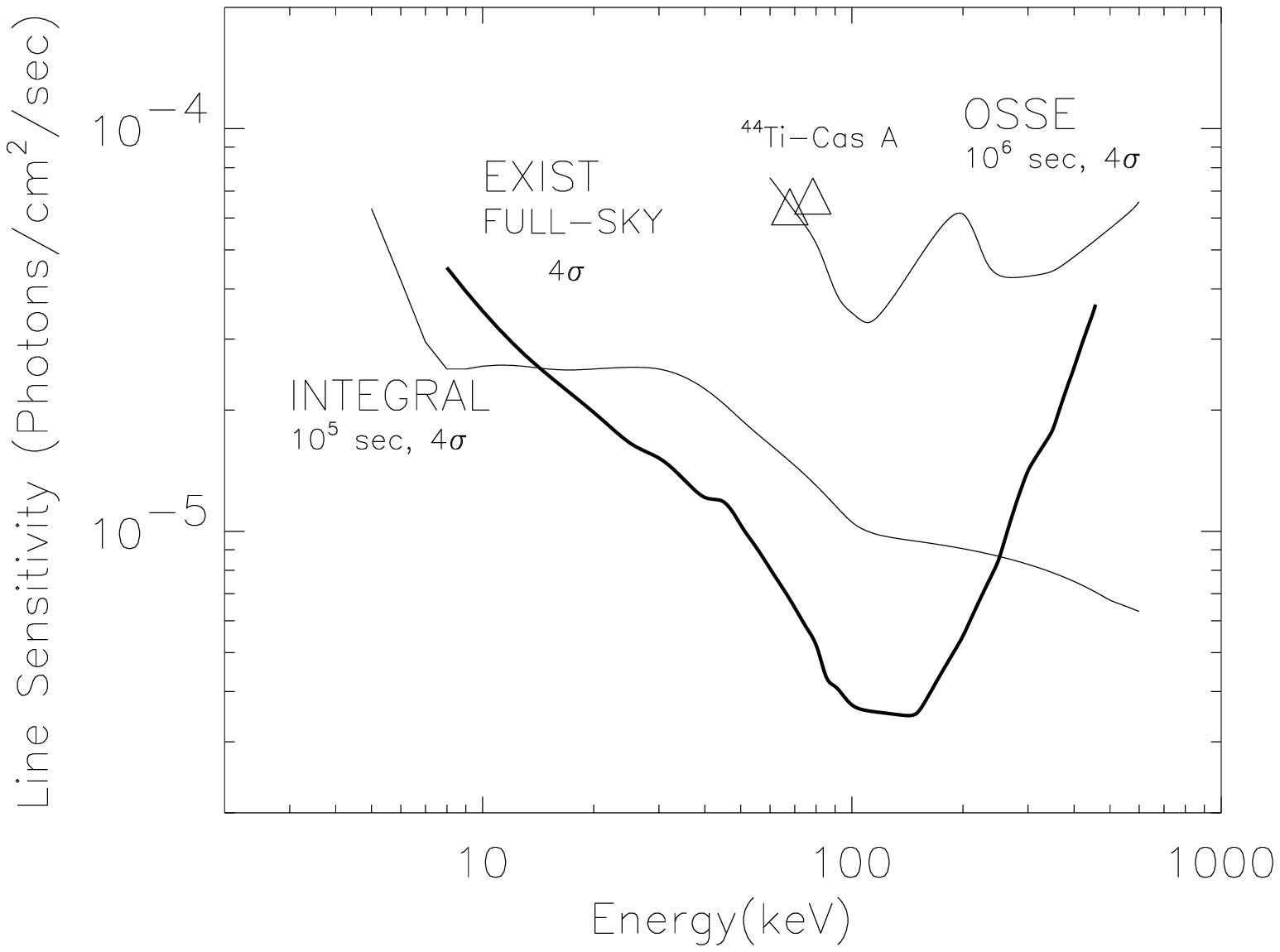,height= 3.2in,width=4.5in,angle=0}

\vspace*{-0.1in}

\ni
Fig. 2. Sensitivity of EXIST (MIDEX mission) for continuum (top)
and narrow lines (bottom).

\newpage
\ni
SCIENTIFIC OBJECTIVES FOR EXIST-LITE
\medskip

\ni
The scientific objectives for the balloon-borne mission, 
EXIST-LITE, would be essentially the same as for the full MIDEX 
mision. Here we list these objectives, 
with a brief discussion of the differences that would apply to 
the 100dLDB type of mission. In general, the balloon mission 
can obtain significantly greater exposure times (particularly 
for high-$\delta$ sources) on intermediate timescales (hours-weeks). 
The entire survey is conducted (for half the sky) in a single 
$\sim$100day flight (vs. 9 months - 2 years for the MIDEX mission). 
However on the very shortest timescales (\la15  min) appropriate 
to an un-occulted observation in a single satellite orbit, 
the MIDEX mission is appreciably (factor of $\sim$2-3) more 
sensitive since it is not affected by  absorption and scattering of 
any overlying atmosphere. Thus the 100dLDB vs. MIDEX versions of 
EXIST can have complementary capabilities and, accordingly, primary 
objectives.
\medskip

\ni
\underline{Hard x-ray spectra and variability of AGNs }\\

\ni
EXIST will have an all-sky sensitivity some 10$\times$ better than
that needed to detect the ``typical'' Seyferts seen with OSSE. 
More than 1000 AGN should be detected in the all sky survey and at least 
$\sim$100-300 could be detected in a 100dLDB EXIST-LITE mission. 
EXIST-LITE has the required sensitivity in 
the poorly explored 30-100 keV band to detect all known AGN detected 
with Ginga or with the Einstein slew survey.  
\medskip

\ni
\underline{Test of halo models and precise positions for Gamma-ray Bursts}\\

\ni
EXIST would have a GRB sensitivity approximately 20$\times$ that of
BATSE so that a 2 month pointed exposure could 
both detect and map a halo in the Andromeda
galaxy (M31) should one exist. EXIST-LITE could achieve a comparable 
exposure if M31 pointings were carried 
out for the $\sim$8 hours of visibility each day although the 
burst sensitivity would likely be reduced (for comparable 
backgrounds) by the factor of $\sim$2 mentioned above. For the
observed GRB logN-logS relation, EXIST-LITE should detect GRBs 
at about 1/2 the rate, or $\sim$0.5/day, as EXIST (or BATSE, with its 
much larger FOV but reduced sensitivity). GRBs will be
located to \la1-5\arcmin positions, thereby providing definitive tests
of repeaters. Bright burst positions and spectra could be brought down
in real time for automated followup searches.
\medskip

\ni
\underline{Studies of black hole and neutron star compact binaries and
transients}\\

\ni
EXIST-LITE surveys 50\% of the sky each day (i.e. the full northern 
hemisphere or sky with $\delta$ \ga 0\deg for a 100dLDB flight at 
mid-latitude of 45\deg) and would achieve 
continuous integration times of  \ga3-12 hours for each source 
each day. Thus studies of compact objects on a wide variety of 
timescales are possible throughout the Galaxy. A deep galactic survey
for transients, black hole binaries and pulsars will allow the
relative populations of black holes in the Galaxy to be constrained. 
Relatively short-duration ($\sim$1 day) transients are particularly 
favorable for discovery and study by EXIST-LITE. 
A much larger population of fainter sources (e.g. black holes with 
hard spectra) could also be detected. A galactic plane survey and 
monitoring for even just a single 100dLDB balloon flight would be highly 
complementary to the planned  galactic plane 
survey with INTEGRAL, which as a series of discrete pointings requires 
$\sim$2 weeks to cover the galactic plane. Since EXIST-LITE only 
observes half the sky in a full flight, it is not able to 
perform the same all-sky monitor function as the full MIDEX mission 
and is, again, complementary. 
\medskip

\ni
\underline{Monitoring and Study of X-ray Pulsars}\\

\ni
The measurement and monitoring of spin periods and luminosity/spectra 
of a large sample of accretion-powered pulsars would greatly extend the 
studies so successfully carried out with BATSE in the $\sim$20-100 keV 
band. With the greatly increased sensitivity and resolution (both spectral 
and spatial), and assuming a logN-logS with slope $\sim$1 for the galactic
population of accretion-powered pulsars, 
EXIST could extend the BATSE sample by a factor of
$\sim$10 or, at the very least, to the entire sample of known 
accretion pulsars. On a 100dLDB mission, EXIST-LITE would 
provide more densely sampled coverage for some sources. In addition, 
the high spectral resolution afforded by the CZT detector array for 
EXIST (or EXIST-LITE) would allow a high sensitivity study of
cyclotron features in pulsar spectra. 
\medskip

\ni
\underline{Emission line surveys: hidden supernovae via $^{44}$Ti
emission and 511 keV sources}\\

\ni
The array of Cd-Zn-Te imaging detectors proposed for EXIST or
EXIST-LITE achieves high spectral resolution (e.g. $\sim$5\% at 60 keV). 
Thus emission line surveys can be conducted. 
The decay of $^{44}$Ti (lines at 68 and 78 keV) with long (68 y)
halflife allows a search for the long-sought population of
obscured supernovae in the galactic plane at sensitivities 
significantly better than the possible detection of Cas-A (cf. Figure
2). These objects would likely appear as discrete (unresolved) 
emission line sources. Similarly, 511 keV emission from black hole 
binaries (or AGN) can be searched for (e.g. in transient outbursts),  
and the diffuse galactic 511 keV emission imaged with sensitivity 
comparable to OSSE (cf. Figure 2). 
\medskip

\ni
\underline{Study of the diffuse hard x-ray background}\\

\ni
The spectra of a significant sample of AGN will test the AGN 
origin of the diffuse background  for the poorly
explored hard x-ray band. Because the background measured by
the EXIST detectors below 100 keV is dominated by the cosmic diffuse
spectrum, its isotropy and fluctuation spectrum can be studied 
with much higher sensitivity than before. This will be more difficult 
to do with EXIST-LITE because of the additional contribution of 
atmospheric background, which peaks in the $\sim$60-100 keV band. 
However this can be modelled and the diffuse background 
isotropy and spectrum could be studied in a series of 100dLDB flights. 

\bigskip

\ni
DETECTOR AND TELESCOPE CONCEPT
\medskip

\ni
\underline{Detector Concept}
\medskip

\ni
EXIST-LITE would incorporate initially one, but ultimately two, large area 
(c. 2500 cm$^2$ each) arrays of Cd-Zn-Te (CZT) detectors as 
proposed for EXIST. Although the detector and telescope design for 
the full MIDEX version of EXIST may differ, here we describe a detector 
and telescope concept for EXIST-LITE that is based on our present 
development of a prototype CZT imaging detector, EXITE3.
\medskip

\ni
The EXIST-LITE detector concept incorporates 2.5mm pixellation on 12 mm (square)
CZT detector elements (5-10mm thick), arrayed in a modular configuration. The
individual 4 $\times$ 4 pixel detector elements are each read out by 
a preamp-shaper-multiplexer ASIC readout circuit with very low
power dissipation (c. 1mW/channel). The 16-channel ASIC would include 
a comparator to examine the multiplexed outputs and output the 1, 4, or 
16 peak channels (configured on command) so that 
multi-site detection (e.g. Compton events and internal background rejection) 
could be accomplished. The ASIC would be packaged  with a 
ceramic/PCB carrier on top (CZT side) with the 4 $\times$ 4 pixel grid of 
contact pins making  contact (with silver epoxy or pogo pins) with the CZT 
pixel array and conductive traces directly connected to the ASIC 
inputs. The bottom (output) side of the ASIC would be socketed with a 
standard multi-pin connector supplying operating voltages, grounds, 
control lines and output (multiplexed) shaped outputs. The CZT-ASIC 
package would be mechanically coupled and locked in a thin-walled 
(0.3 mm) copper square tube (12.5mm $\times$ 12.5mm $\times$ 10mm) 
epoxied to a closed and insulating (e.g. Kevlar) 
end cap which provides mechanical rigidity and light insulation.  
A thin (1mm) conductive rubber washer epoxied to the central 10mm 
diameter of the end cap is used to supply the negative bias voltage 
(-500V) to the radiation side of the CZT detector. The washer is 
connected with a small feed-through pin in the corner of the end cap 
to an external (top) bias distribution network for the contiguous CZT 
detectors in the array.
\medskip

\ni
This packaging concept thus provides for complete modularity of 
individual detector and ASIC for plug in (and easy replacement) 
on a purely analogue mother board. The individual CZT-ASIC 
modules are sized for (relatively) easy fabrication and acquisition 
of CZT (i.e. 12mm x 12mm x 5mm crystals), noise isolation 
and thermal coupling to the mechanical frame and BGO collimator 
mounted above. 
\medskip

\ni
For the 45\deg (FWHM) field of view desired for EXIST-LITE, 
the collimator blade height/spacing ratio is tan (45\deg) =  1, and 
for the collimator (assumed to be BGO; see below) to have \ga50\% collimation 
efficiency at all energies below 600 keV, it must have wall thickness of 
3mm (i.e. \ga5mm in collimation). For the collimator to also provide 
active shielding for the n-$\gamma$ background expected in the CZT 
(see below), it must be \ga5-10mm thick for the packaging discussed 
below. Thus the collimator pitch must be (significantly) larger than 
the CZT-ASIC module size (12mm pitch) or it will block an 
unacceptably large fraction of the open area. We choose a collimator 
pitch of 10cm, since this will yield a convenient Basic Detector 
Module (BDM) size of 8 x 8 CZT-ASIC modules with total area of 100 cm$^2$. 
This leads, then, to a choice of the following concept for the detector 
digitization and overall packaging: the full detector is tiled with 
100cm$^2$ self-contained 
detector modules (BDM), each with surrounding (4-sided) 
7 mm thick BGO collimator (10cm high) providing 
external shielding, and isolated ADC and interface to the detector 
data bus. Behind the analogue mother-board (10cm x 10cm) for the 
BDM would then be a 2cm thick BGO anti-coincidence shield 
which is optically coupled to the 4-sided BGO collimator shield 
above. Because of the close packing of the CZT-ASIC and motherboard, 
it would then be only $\sim$2cm behind the rear of the CZT. 
All analogue signals from the motherboard would be routed 
out through a half-hole in one corner of the shield, which 
would (when the full detector is tiled) require then only 
one shield feed through for 4 contiguous BDMs. The BGO 
shield would itself be readout with 2 avalanche photo-diodes 
(APDs) optically coupled on opposite sides of its bottom (digital) side. 
Each of the two complete EXIST-LITE detectors would thus consist of 
an array of 5 $\times$ 5 or 
25 such BDMs. These modules would be close-packed 
and mounted in a common frame. Although the 
collimator shields will require a net gap of $\sim$1.5 cm between 
each of the 5 BDMs across the detector array, this does not affect 
image reconstruction but only makes the detector $\sim$8 cm larger on a side. 

\medskip

\ni
By incorporating an active collimator of BGO, the background 
rejection in each BDM is expected to be optimum: the n-$\gamma$ 
activation background induced in the Cd-Zn-Te detector by 
the interaction of local thermal neutrons in the Cd will be more 
efficiently rejected since the collimator now actively shields $\sim$90\% 
of the forward hemisphere (vs. having the active BGO shield only 
shield the rear hemisphere of the CZT detector array as originally 
considered for EXIST (cf. Grindlay et al 1995). It is likely 
that this modular construction and active collimation for the 
high energy FOV would be favorable for the satellite version of 
EXIST also.
\medskip

\ni
This 10cm $\times$ 10cm packaging concept 
for the BDM and detector-assembly allows 
for the incremental development of the detector, as needed for a 
balloon-flight development program where cost constraints are  
significant. A key difference from the satellite EXIST mission 
is that the low energy collimator could  be eliminated 
since it was primarily intended to reduce the cosmic diffuse and 
point source background below 30 keV by restricting the low energy 
FOV (below $\sim$30-40 keV) to approximately 10\% of the solid 
angle of the high energy FOV. Since EXIST-LITE would have 
a low energy cutoff of 25-30 keV imposed by the overlying residual 
atmosphere (effectively $\sim$4 g/cm$^2$) at its expected altitude appropriate 
for a LDB flight, this low energy collimation is probably not needed.  
\medskip

\ni
The development of a single BDM could occur on a 
timescale of  $\sim$1-2 years and would allow a prototype EXIST-LITE 
imager, EXITE3,  to be demonstrated (complete with coded mask; see below) 
as a piggy-back on the current EXITE2 telescope and pointing 
system.
\medskip

\ni
\underline{Telescope Concept}
\medskip

\ni
The coded aperture telescopes for either a single EXITE3 prototype module 
or the eventual EXIST-LITE system 
can be relatively compact design with coded mask at focal length 1.43m and mask
pixel size 5mm. This yields an imaging resolution of 12\arcmin, 
which is appropriate to resolve even the most crowded galactic bulge 
fields at the high sensitivity expected. In order to cover the 
full 45\deg field of view, 
the EXITE3 prototype would have a URA mask of approximate dimensions 
1.3m $\times$ 1.3m and format 257 $\times$ 255 to fully image the 45\deg 
FWHM field of view. However, since the BGO collimator on each BDM 
segment of the detector array would produce partial coding for 
sources off-axis, the mask must be either smaller format and repeated 
(e.g. 4 contiguous 127 $\times$ 129 masks, 
leading to ambiguous source positions) or random. 
A random mask would be, in any case, as effective as a URA 
of such large format. 
The complete EXIST-LITE telescope need 
only have a mask some 0.5m larger in each dimension or $\sim$
1.7m. Since 
the coded mask should not collimate the image significantly, 
the mask thickness is 
restricted to be \la5mm, which (for Ta mask elements) restricts its 
upper energy limit to be \la600 keV for partial shadowing. The full 
dimensions of the mask, and thus its mass ($\sim$150 kg for the 
full mask) can of course be reduced by decreasing 
the focal length by either  degrading 
the angular resolution or detector oversampling  (or both).

\newpage

\ni
CZT DETECTOR AND ARRAY STUDIES
\medskip

\ni
As part of the effort to both conduct the EXIST Mission Concept study 
and optimize the design for a future MIDEX proposal as well 
as EXIST-LITE, we are 
conducting a variety of studies of CZT detectors and array technologies 
at CfA and in collaboration with both other EXIST Team members and 
industry.
\medskip

\ni
\underline{Balloon Flight Tests of Backgrounds and Shielding Efficiency}
\medskip

\ni
A critical area of concern for the use of CZT detectors in space is 
the possibly  high levels of internal background they might experience 
due to the large neutron cross section(s) for Cd, which result in 
prompt gamma-ray decays. Balloon flight tests of single 
isolated CZT detectors by the GSFC and Caltech groups in 
May and September-October, 1995, suggested disturbingly 
large in-flight backgrounds compared to those expected for 
similar scintillation detectors (e.g. Parsons et al 1996). However 
the GSFC measurement of a marked reduction in background with 
an external anti-coincidence shield (NaI) suggested this could 
be effectively reduced by suitable active shielding. 
\medskip

\ni
In collaboration with Caltech, we have assembled 
a flight unit to test the prompt anti-coincidence shielding 
efficiency of a planar BGO shield immediately behind the 
CZT detector plane, as proposed for EXIST. The 
BGO (75mm diameter $\times$ 75mm thick, 
and supplied by JPL) is centered below a single element CZT detector 
(10mm $\times$ 10mm $\times$ 2mm, and supplied by eV Products 
to Caltech). The detector-shield and preamp are mounted in a 
pressure vessel and shielded with a 1.8mm thick Pb + 0.8mm thick Sn 
and 1.2mm Cu graded shield to simulate the approximate grammage 
of the passive collimator 
in front of an EXIST detector. The raw CZT and BGO (shield) detector 
preamp outputs are interfaced to shaping amps and digital (discriminator 
and 12 bit ADC) electronics built at CfA to interface to the 
flight computer and data system 
for the EXITE2 balloon-borne telescope. Unfortunately, due to two 
successive campaigns (May and September-October, 1996) of  high 
surface winds and no launch opportunities, this experiment is still 
awaiting a Ft. Sumner launch (May 1997). 
\medskip

\ni
\underline{Balloon Flight Measure of Neutron Backgrounds}
\medskip

\ni
In order to fully calibrate the CZT background and shielding 
experiment so that balloon results may be extrapolated to 
the full space environment, a simultaneous measure of the neutron flux 
experienced by the detector is desirable. The atmospheric neutron 
fluxes  as tabulated by Armstrong (1973) are 
sufficiently uncertain (probably by a factor of \ga2) that 
we shall attempt to measure the flux by a simple passive experiment: 
an array (7 $\times$ 6) of gold foils (each $\sim$6cm$^2$) mounted 
on top of the gondola in which the n-$\gamma$ reaction  
\medskip

Au-197(n,$\gamma$)~~$\rightarrow$~~~Au-198(,e)Hg-198
\medskip

\noindent
will be measured after the flight by observing the resulting 
412 keV decay $\gamma$-ray (2.7d halflife) with a low 
background Ge spectrometer at JPL (by L. Varnell).  This 
experiment, conducted in collaboration with G. Skinner 
and L. Varnell, is also awaiting the Ft. Sumner launch.
\medskip

\ni
\underline{Spatial Uniformity of CZT}
\medskip

\ni
Pixellated CZT detector arrays, as proposed for 
the full EXIST mission or EXIST-LITE, will require 
relatively uniform response across both the projected surface area 
and depth of the detector elements. Non-uniformities of detector 
response can be calibrated out (by flat fielding) but will be simplified 
to the extent the detectors are uniform (and will be less of a 
problem with the relatively large pixel detectors for EXIST 
than with small pixel CZT imagers for focussing optics). 
We have conducted a program of  mapping the spectral 
response of single detectors and comparing the observed 
variations with IR micrographs (obtained at eV Products) 
of  the detector to correlate 
spectral response with grain boundaries and inclusions in the 
detector. Spectra (Am-241) obtained in a 3 $\times$ 3 raster 
scan of a 0.5mm beam across  a 4mm $\times$ 4mm $\times$ 3mm CZT detector 
show variation in spectral response which correlates  with the grain boundaries  
as well as inclusions and precipitates. Results were reported by us 
at the NASA-SEU Technology Workshop (December, 1996).  
\medskip

\ni
\underline{Development of PIN Readouts for CZT Detectors}
\medskip

\ni
CZT detectors are conventionally fabricated with metal (gold) 
contacts deposited directly on the CZT crystal. These 
metal-semiconductor-metal (M-S-M) detectors are of course the subject 
of intense development and are baselined for EXIST. However, they 
suffer from limitations of charge collection efficiency (although
these are at least partially overcome with ``small pixel'' electrodes; cf. 
Barrett et al 1995) and poor ohmic contacts. Several groups, most 
recently SBRC (Hamilton et al 1996) have investigated alternative 
readouts incorporating P-I-N junctions. The Spire Corp. (Bedford, MA) 
has developed a  new method for fabrication of P-I-N electrodes on 
CZT by using CdS(p-type) and ZnTe(n-type) layers deposited by thermal 
evaporation on both high pressure Bridgman (HPB) CZT crystals (from eV
Products) as well as lower cost vertical Bridgman (VB) crystals (from 
Cleveland Crystals), and the results appear very encouraging. 
At CfA we are testing these P-I-N readout CZT detectors which offer 
advantages of improved charge collection and ease of fabrication for 
their use as thick detectors. We are working with Spire to fabricate 
a 4 $\times$ 4 array P-I-N detector (on a 10mm $\times$ 10mm 
$\times$ 5mm CZT substrate) for balloon flight tests of background and 
uniformity of response.
\medskip

\ni
\underline{Development of Thick CZT Detector Array Readouts}
\medskip

\ni
Thick detectors (5mm or greater) as desired for EXIST pose 
special challenges for the optimum design of the detector and 
readout. In particular, the electric field configuration 
needed for the small pixel effect (Barrett et al 1995) must be 
carefully considered, and the effects of charge diffusion and 
spreading become more important. We are exploring these effects 
in collaboration with the RMD Corporation (Watertown, MA), who 
have developed evaporative mask techniques for array fabrication 
and have built a prototype array (3 x 3; 1mm pixels) using 4mm thick 
CdTe grown by RMD. Results for Co-57 show high 
photopeak efficiency and uniformity. In collaboration with RMD, we have 
now extended this to CZT: RMD has just completed fabrication 
of a prototype 4 $\times$ 4 array M-S-M detector  (on a 10mm $\times$ 10mm 
$\times$ 5mm CZT substrate)  and initial results appear very 
promising and will be reported in Shah et al (1997). 
At CfA, we shall test fully this array for its small-pixel effect 
properties. We are also integrating the array with an ASIC readout, 
using a prototype 16-channel preamp/shaper ASIC developed 
by IDE Corp. The same ASIC readout system will allow comparative tests
of the MSM array detectors from RMD and Spire, and the PIN array 
being developed by Spire.  Initial results will be presented in a 
forthcoming paper by Bloser et al (1997). 
\bigskip

\ni
ACKNOWLEDGEMENTS
\medskip

\ni
The EXIST Concept was developed by the author in close collaboration 
with T. Prince, F. Harrison, C. Hailey, N. Gehrels, B. Ramsey, M. Weisskopf, 
G. Skinner and P. Ubertini, all of whom are gratefully acknowledged. The 
EXITE3 and CZT development is being done with P. Bloser, Y. Chou, 
G. Monnelly, T. Narita and S. Romaine, all of whom are also thanked. 
This work was supported in part by NASA grants  NAGW-624 and NAG5-5103.

\bigskip

\bigskip

\ni
REFERENCES
\medskip

\ref{Armstrong, T.W., Chandler, K.C. and Barish, J., 
Calculations of Neutron Flux Spectra Induced in the Earth's Atmosphere
by Galactic Cosmic Rays, {\it JGR}, {\bf
78} (16), 2715 (1973).}
\ref{Barrett, H., Eskin, J. and Barber, H, Charge Transport in 
Arrays of Semiconductor Gamma-Ray Detectors, 
{\it Phys. Rev. Letters}, {\bf 75}, 156 (1995).}
\ref{Bloser, P., Grindlay, J., Narita, T. {\it et al}, in preparation 
(1997).}
\ref{Covault, C., {\it Ph.D. Thesis, Harvard University} (1991).}
\ref{Caroli, E., Stephen, J., DiCocco, G., Natalucci, L. 
and Spizzichino, A., Coded Aperture Imaging in X- and Gamma-Ray 
Astronomy {\it Space Sci. Rev.}, {\bf 45}, 349 (1987).}
\ref{Grindlay, J., Prince, T., Gehrels, N.., Tueller, J., Hailey, C. 
{\it et al}, Energetic X-ray Imaging Survey Telescope (EXIST), 
{\it Proc. SPIE}, {\bf 2518}, 202 (1995).}
\ref{Grindlay, J. Prince, T., Gehrels, N., Hailey, C. 
Harrison, F.{\it et al}., 
Proposed (to) EXIST: Hard X-ray Imaging All Sky Monitor, {\it
All Sky X-ray Observations in the Next Decade}, Proc. Workshop in
Riken, Waco, Japan, March 1997, 
(M. Matsuoka and N. Kawai, eds.), in press.}
\ref{Hamilton, W., Rhiger, D., Sen, S., Kalisher, M., Chapman, G. 
and Millis, R., HgCdTe/CdZnTe P-I-N High Energy Photon 
Detectors, {\it Jour. Elec. Mat.}, {\bf 25}, 1286 (1996).}
\ref{Levine, A., Lang, F., Lewin, W., Primini, F., Dobson, C. \etal, 
The HEAO-1 A-4 Catalog of High Energy X-ray Sources, 
{\it ApJ Suppl.}, {\bf 54}, 581 (1984).}
\ref{Lum, K., Manandhar, R., Eikenberry, S., Krockenberger, M. 
and Grindlay, J., Initial Performance of the EXITE2 Imaging Phoswich 
Detector/Telescope for Hard X-ray Astronomy, {\it IEEE Trans. Nucl. 
Sci.}, {\bf NS-41}, 1354 (1994).}
\ref{McConnell, M., Dunphy, P., Forrest, D., Chupp, E. and Owens, A., 
Gamma-ray Observations of the Crab Region Using a Coded Aperture 
Telescope, {\it Ap.J.}, {\bf 321}, 543 (1987).}
\ref{Parsons, A., Barthelmy, S., Bartlett, L., Birsa, F., Gehrels, N. 
\etal, CdZnTe Background Measurements at Balloon Altitudes, 
{\it Proc. SPIE}, vol. 2806, 432 (1996).}
\ref{Paul, J., Mandrou, P. and Ballet, J., SIGMA: The Hard X-ray and 
Soft Gamma-Ray Telescope on board the GRANAT Space 
Observatory, {\it Adv. Sp. Res.}, {\bf 11} (8), 289 (1991).}
\ref{Shah, K., Cirignano, L., Klugerman, M., Dmitreyev, Y., 
Grindlay, J. \etal, Multi-Element CdZnTe Detectors for Gamma-Ray
Detection and Imaging, 
{\it IEEE Trans. Nucl. Sci.}, submitted (1997).}

(and see additional references given in Table 1).

\end{document}